\documentclass[pra,aps,nopacs,nokeys,superscriptaddress,twocolumn,twoside]{revtex4}

\usepackage{graphicx,epic,eepic,epsfig,amsmath,latexsym,amssymb,verbatim,revsymb,color}

\usepackage{theorem}

\def\squareforqed{\hbox{\rlap{$\sqcap$}$\sqcup$}}
\def\qed{\ifmmode\squareforqed\else{\unskip\nobreak\hfil
\penalty50\hskip1em\null\nobreak\hfil\squareforqed
\parfillskip=0pt\finalhyphendemerits=0\endgraf}\fi}
\def\endenv{\ifmmode\;\else{\unskip\nobreak\hfil
\penalty50\hskip1em\null\nobreak\hfil\;
\parfillskip=0pt\finalhyphendemerits=0\endgraf}\fi}

\newlength{\blank}
\newenvironment{proof}[1][{\hspace{-\blank}}]{{\noindent{\bf Proof~{#1}.\ }}}{\hfill\qed\vskip 0.5\baselineskip}

\mathchardef\ordinarycolon\mathcode`\: \mathcode`\:=\string"8000
\def\vcentcolon{\mathrel{\mathop\ordinarycolon}}
\begingroup \catcode`\:=\active
  \lowercase{\endgroup
  \let :\vcentcolon
  }

\newcommand{\nc}{\newcommand}
\nc{\rnc}{\renewcommand} \nc{\beq}{\begin{equation}} \nc{\eeq}{{\end{equation}}} \nc{\beqa}{\begin{eqnarray}}
\nc{\eeqa}{\end{eqnarray}} \nc{\lbar}[1]{\overline{#1}} \nc{\bra}[1]{\langle#1|} \nc{\ket}[1]{|#1\rangle}
\nc{\ketbra}[2]{|#1\rangle\!\langle#2|} \nc{\braket}[2]{\langle#1|#2\rangle} \nc{\proj}[1]{| #1\rangle\!\langle
#1 |} \nc{\avg}[1]{\langle#1\rangle}
\nc{\Rank}{\operatorname{Rank}} \nc{\smfrac}[2]{\mbox{$\frac{#1}{#2}$}} \nc{\tr}{\operatorname{Tr}}
\nc{\ox}{\otimes} \nc{\dg}{\dagger} \nc{\dn}{\downarrow} \nc{\cA}{{\cal A}} \nc{\cB}{{\cal B}} \nc{\cC}{{\cal
C}} \nc{\cD}{{\cal D}} \nc{\cE}{{\cal E}} \nc{\cF}{{\cal F}} \nc{\cG}{{\cal G}} \nc{\cH}{{\cal H}}
\nc{\cI}{{\cal I}} \nc{\cJ}{{\cal J}} \nc{\cK}{{\cal K}} \nc{\cL}{{\cal L}} \nc{\cM}{{\cal M}} \nc{\cN}{{\cal
N}} \nc{\cO}{{\cal O}} \nc{\cP}{{\cal P}} \nc{\cR}{{\cal R}} \nc{\cS}{{\cal S}} \nc{\cT}{{\cal T}}
\nc{\cX}{{\cal X}} \nc{\cZ}{{\cal Z}} \nc{\csupp}{{\operatorname{csupp}}} \nc{\qsupp}{{\operatorname{qsupp}}}
\nc{\var}{{\operatorname{var}}} \nc{\rar}{\rightarrow} \nc{\lrar}{\longrightarrow}
\nc{\polylog}{{\operatorname{polylog}}} \nc{\1}{{\openone}} \nc{\wt}{{\operatorname{wt}}}

\def\a{\alpha}
\def\b{\beta}

\nc{\RR}{{{\mathbb R}}} \nc{\CC}{{{\mathbb C}}} \nc{\FF}{{{\mathbb F}}} \nc{\NN}{{{\mathbb N}}}
\nc{\ZZ}{{{\mathbb Z}}} \nc{\PP}{{{\mathbb P}}} \nc{\QQ}{{{\mathbb Q}}} \nc{\UU}{{{\mathbb U}}}
\nc{\EE}{{{\mathbb E}}} \nc{\id}{{\operatorname{id}}}

\nc{\CHSH}{{\operatorname{CHSH}}}

\nc{\be}{\begin{equation}} \nc{\ee}{{\end{equation}}} \nc{\bea}{\begin{eqnarray}} \nc{\eea}{\end{eqnarray}}
\nc{\<}{\langle} \rnc{\>}{\rangle} \nc{\Hom}[2]{\mbox{Hom}(\CC^{#1},\CC^{#2})} \nc{\rU}{\mbox{U}}

\nc{\ob}[1]{#1}

\newcommand{\av}[1]{\left\langle #1 \right\rangle}

\newcommand{\eff}{\mathrm{\,eff}}

\newcommand{\swap}{\mathbb{S}}

\begin{document}

\title{Quantum mechanical evolution towards thermal equilibrium}

\author{Noah Linden}
\affiliation{Department of Mathematics, University of Bristol, University Walk, Bristol BS8 1TW, U.K.}

\author{Sandu Popescu}
\affiliation{H.H.Wills Physics Laboratory, University of Bristol, Tyndall Avenue, Bristol BS8 1TL, U.K.}
\affiliation{Hewlett-Packard Laboratories, Stoke Gifford, Bristol BS12 6QZ, U.K.}

\author{Anthony J. Short}
\affiliation{Department of Applied Mathematics and Theoretical Physics, University of Cambridge, Centre for
Mathematical Sciences, Wilberforce Road, Cambridge CB3 0WA, U.K.}

\author{Andreas Winter}
\affiliation{Department of Mathematics, University of Bristol, University Walk, Bristol BS8 1TW, U.K.}
\affiliation{Centre for Quantum Technologies, National University of Singapore, 2 Science Drive 3, Singapore
117542}

\date{12 December 2008}

\begin{abstract}
The circumstances under which a system reaches thermal equilibrium, and how to derive this from basic dynamical
laws, has been a major question from the very beginning of thermodynamics and statistical mechanics. Despite
considerable progress, it remains an open problem. Motivated by this issue, we address the more general question
of equilibration. We prove, with virtually full generality, that reaching equilibrium is a universal property of
quantum systems: Almost any subsystem in interaction with a large enough bath will reach an equilibrium state
and remain close to it for almost all times. We also prove several general results about other aspects of
thermalisation besides equilibration, for example, that the equilibrium state does not depend on the detailed
micro-state of the bath.
\end{abstract}

\maketitle

Leave your hot cup of coffee or cold beer alone for a while and they soon lose their appeal - the coffee cools
down and the beer warms up and they both reach room temperature. And it is not only coffee and beer - reaching
thermal equilibrium is a ubiquitous phenomenon: everything does it. Thermalization is one of the most
fundamental facts of nature.

But how exactly does thermalization occur? How can one derive the existence of this phenomenon from the basic
dynamical laws of nature (such as Newton's or Schr\"odinger's equations)? These have been open questions since
the very beginning of statistical mechanics more than a century and a half ago.

One - but by no means the only - stumbling block has been the fact that the basic postulates of statistical
mechanics rely on subjective lack of knowledge and ensemble averages, which is very controversial as a physical
principle. Recently however there has been significant progress: it was realized that ensemble averages and
subjective ignorance are not needed, because {\it individual} quantum states of systems can exhibit statistical
properties. This is a purely quantum phenomenon, and the key is entanglement, which leads to {\it objective}
lack of knowledge. Namely, in quantum mechanics even when we have complete knowledge of the state of a system,
i.e. it is in a pure state and has zero entropy, the state of a subsystem may be mixed and have non-zero
entropy. In this situation we cannot describe the subsystem by any particular pure state, and for all purposes
it behaves as if we have a lack of knowledge about what its pure state is (i.e. it behaves as a probability
distribution over pure states). This is in stark contrast to classical physics where complete knowledge of the
state of the whole system implies complete knowledge of the state of any subsystem, and hence probabilities can
only arise as purely subjective lack of knowledge (i.e. the subsystem has a well-defined state only we don't
know what that is).

This approach has become a very fruitful direction of research in recent
years~\cite{Mahler:etal,Lebowitz,PopescuShortWinter}, see also important earlier
work~\cite{bocchieri:loinger,lloyd,Deutsch-Srednicki,Tasaki} and numerical studies~\cite{Rigol-etal}.

Most notably, it was shown in \cite{Mahler:etal,Lebowitz,PopescuShortWinter} that almost all (pure) states of a
large system are such that any small subsystem is in a canonical state, as if it were the result of averaging
(with equal probability) over all possible states respecting the desired macroscopic constraints. However, the
above result is limited in that it refers to the state at a given time, and deals only with the case of generic
states. But what we are most interested in is not the situation at one given moment of time but in the time
evolution - in particular, under what circumstances systems reach equilibrium, and how much they fluctuate.
Furthermore, the states of most interest, namely those initially far from equilibrium, are non-generic. These
dynamical aspects are the subject of the present paper.

Thermalisation seems a very straightforward process - put a system in contact with a large enough thermal bath
and the system will reach equilibrium at the same temperature. Closer analysis reveals however that the process
of thermalisation actually contains many different aspects and we can decompose it into the following elements.
They constitute a rough roadmap of what one has to show in order to demonstrate that a particular physical
system thermalises.

\begin{enumerate}
\item {\it Equilibration}. We say that a system equilibrates if its state evolves towards some
particular state (in general mixed) and remains in that state (or close to it) for almost all times. As far as
equilibration is concerned, it is irrelevant what the equilibrium state actually is. In particular, it need not
be a thermal state (such as a Boltzmannian distribution) and indeed may depend on the initial state of the
subsystem and/or on the initial state of the rest of the system (i.e. bath) in an arbitrary way.

\item{\it Bath  state independence}. The equilibrium state of the system should not
depend on the precise initial state of the bath. That is, when defining a bath, we should only specify its
macroscopic parameters, e.g. its temperature; when the system reaches equilibrium, the equilibrium state should
depend only on the temperature of the bath.

\item{\it Subsystem state independence}. If the subsystem is small
compared to the bath, the equilibrium state of the subsystem should be independent of its initial state.

\item {\it Boltzmann form of the equilibrium state.} Under certain
additional conditions on the Hamiltonian (especially the interaction term) and on the initial state, the
equilibrium state of the subsystem can be written in the familiar Boltzmannian form $\rho_S =
\frac{1}{Z}\exp\left(-\frac{H_S}{k_B T}\right)$.
\end{enumerate}

Realizing that thermalization can be decomposed in this way has major consequences. Firstly, it allows us to
address each aspect separately. Secondly, and more important, it allows us to greatly expand the scope of our
study. Indeed, we will consider equilibration as a general quantum phenomenon that may occur in situations other
than those usually associated with thermalization. In particular we need not restrict ourselves to: standard
thermal baths (that are described by a given temperature or restricted energy range), weak or short range
interactions between the system and the bath, Boltzmannian distributions, situations in which energy is an
extensive quantity, etc. Furthermore, we can consider situations in which the systems do not reach equilibrium,
and prove results about the bath or subsystem independence properties of the time-averaged state.

In this paper, with very weak assumptions, we prove the first two elements above - equilibration and bath state
independence. That is, with virtually full generality, we prove that reaching equilibrium is a universal
property of quantum systems, and that the equilibrium state does not depend on the precise details of the bath
state. After introducing the setup and basic definitions in the next section, we go through the above programme
step-by-step.

\section{Setup and Definitions}
\label{sec:setup} In this section, we describe our general setup, and introduce the basic notations and
definitions that will be used throughout the paper.

{\medskip\noindent \textbf{The system.} We consider a large quantum system, described by a Hilbert space
$\mathcal{H}$. We decompose this system into two parts, a small subsystem $S$, and the rest of the system that
we refer to as the bath $B$. Correspondingly, we decompose the total Hilbert space as $\mathcal{H}
=\mathcal{H}_S \otimes \mathcal{H}_B$, where $\mathcal{H}_S$ and $\mathcal{H}_B$ (of dimensions $d_S$ and $d_B$)
are the Hilbert spaces of the subsystem and bath respectively. If either part is infinite-dimensional, we
introduce a high-energy cut-off to render its dimension finite (and eliminate interaction terms from the
Hamiltonian that would take the state outside the allowed subspace). Note, however, that at this stage we do not
ascribe the subsystem or bath any special properties. The subsystem S could be a single particle, a cluster of
particles next to each other, or a number of isolated particles spread throughout the bath. It may even be
something far more abstract, not necessarily related to particles at all: Any arbitrary decomposition of the
Hilbert space into a tensor product ($\mathcal{H} = \mathcal{H}_S \otimes \mathcal{H}_B$) actually defines a
subsystem and a bath.}

\medskip\noindent
\textbf{The Hamiltonian.}The evolution of the total system is governed by a Hamiltonian
\begin{equation}
  H= \sum_k E_k \proj{E_k},
\end{equation}
where $\ket{E_k}$ is the energy eigenstate with energy $E_k$. Throughout this paper, we consider the Hamiltonian
to be completely general, except for the following constraint: That it has \emph{non-degenerate energy gaps}.

We say that a Hamiltonian has non-degenerate energy gaps if any nonzero difference of eigenenergies
determines the two energy values involved. 
I.e., for any four eigenstates with energies $E_k$, $E_\ell$, $E_m$ and $E_n$,
$E_k - E_\ell = E_m - E_n$ implies $k=\ell$ and $m=n$, or $k=m$ and $\ell=n$.
Note that this also implies that the energy levels are non-degenerate.

{An important physical implication of this assumption is that the Hamiltonian is fully interactive, in the sense
that no matter how we partition the system into a subsystem and bath they interact. }Indeed any non-interacting
Hamiltonian $H=H_S + H_B$ has multiple degenerate energy gaps. Note in particular that for non-interacting
Hamiltonians the energy is a sum of the system energy and bath energy. Any four energies satisfying $E_1 =
E^S_1+E^B_1$, $E_2 = E^S_1+E^B_2$, $E_3 = E^S_2+E^B_1$, and $E_4 = E^S_2+E^B_2$ will lead to a degenerate gap.

We emphasize that the restriction to Hamiltonians that have no degenerate energy gaps is an extremely natural
and weak restriction. Indeed, adding an arbitrarily small random perturbation to any Hamiltonian will remove all
degeneracies (although such changes may take a long time to significantly influence the evolution of a state,
here we are not concerned with the time scales).

{ Note that except for the Hamiltonian having non-degenerate energy gaps, we allow it to be completely general.
In particular, in a system composed of particles, it need not describe nearest-neighbour or bipartite
interactions, but could contain interactions between all particles simultaneously. Hence energy need not be an
(even approximately) extensive quantity, as normally considered in statistical mechanics.}

\medskip\noindent
\textbf{Notation.} We denote by $\ket{\Psi(t)}$ the global pure state of the system (i.e. of the subsystem and
bath together) at time $t$. It is also convenient to write this state as a density matrix $\rho(t) =
\proj{\Psi(t)}$. The state of the subsystem $\rho_S(t)$ is obtained by tracing the global state $\rho(t)$ over
the bath. I.e. $\rho_S(t)=\tr_B \rho(t) $. Similarly the state of the bath at time $t$ is described by
$\rho_B(t)=\tr_S \rho(t) $.


It is useful to define the time-averaged state of the system $\omega$, which is given by
\begin{equation} \label{eqn:notation1}
  \omega = \av{\rho(t)}_t = \lim_{\tau \rightarrow \infty} \frac{1}{\tau} \int_0^{\tau} \rho(t) {\rm d}t.
\end{equation}
Similarly we define $\omega_S$ and $\omega_B$ as the time-averaged states of the system and bath respectively.

It is also convenient to introduce the notion of the \emph{effective dimension} of a (mixed) state $\rho$ by
\begin{equation} \label{eqn:notation2}
  d^{\eff} (\rho) = \frac{1}{\tr(\rho^2)}.
\end{equation}
This tells us, in a certain sense, how many pure states contribute appreciably to the mixture. In particular a
mixture of $n$ orthogonal states with equal probability has effective dimension $n$. Unlike the support of the
density matrix, this notion captures the probabilistic weight of different states in the mixture, and is
continuous.

It will also be important in what follows to consider the distance between two density matrices. There are many
possible ways to define such a distance; here we use a strong and very natural distance,  the trace-distance
$D(\rho_1, \rho_2 )=\frac{1}{2} \tr \sqrt{ ( \rho_1 - \rho_2 )^2 }$. The trace-distance characterises how hard
it is to distinguish two states experimentally (even given perfect measurements). When it is small, the two
states are effectively indistinguishable. More precisely, it is equal to the maximum difference in probability
for any outcome of any measurement performed on the two states. Furthermore, the maximum difference in the
expectation values of any operator $A$ on the two states is the range of the eigenvalues of the operator times
the trace-distance $(a_{max} - a_{min}) D$.

Finally, by ${\Pr}_{\Psi} \{ \cdot \}$ we denote the proportion of states in a Hilbert space having a particular
property, according to the natural Haar measure.

\section{Equilibration}
\label{sec:equalibration} We now come to the central result of our paper: {\it Every pure state of a large
quantum system that is composed of a large number of energy eigenstates and which evolves under any arbitrary
Hamiltonian (with non-degenerate energy gaps) is such that every small subsystem will equilibrate.} That is,
every small subsystem will evolve towards some particular state (in general mixed) and remain in that state (or
close to it) for almost all times.

To understand the requirement that the state contains a large number of energy eigenstates we first note that
this means that there will be a lot of change during the time evolution. Indeed, a single energy eigenstate does
not change at all. (In this trivial sense, every energy eigenstate is already equilibrated. We however are
looking for systems that do not start in equilibrium but evolve towards it.) But why do we need significant
change during the time evolution? In order for equilibration to occur, we need that some information about the
initial state of the subsystem leaves the subsystem. Indeed, suppose we start with a subsystem far from
equilibrium. Then it will evolve through a number of distinct states on its path towards equilibrium. This
implies that the state of the whole system (i.e. subsystem and bath) will also evolve through a number of
distinct states. However, suppose now that the subsystem reaches equilibrium. By definition, this means that its
state doesn't change (significantly) any more. Nevertheless, due to unitarity the state of the whole system
cannot simply stop evolving, and it must continue to go through distinct states at the same rate as initially.
In order for recurrences of the non-equilibrium state of the subsystem to occur very infrequently, the states of
the whole system in which the subsystem is far from equilibrium must be only a small fraction of the total
number of distinct states through which the whole system evolves. Hence the whole system must evolve through
very many states. In order for this to be possible, the state of the total system must be composed of many
energy eigenstates.


We have argued above that for a subsystem initially out of equilibrium to reach equilibrium, the state of the
whole system must go through many distinct states. What we prove in the present paper is that this is also
sufficient. That is:
\begin{quote} {\it Whenever the state of the whole system goes through many distinct states any small
subsystem reaches equilibrium.} \end{quote}

Another way of looking at the time evolution of the state is to investigate what happens to the state of the
bath. As mentioned above, due to unitarity, the state of the whole system must continue evolving at the same
rate even when the state of the subsystem is at equilibrium and doesn't change. The total state can then change
in two ways: through changes in the correlations between the  subsystem and the bath and through changes of the
state of the bath itself. What we prove is:
\begin{quote} {\it Whenever the state of the {\rm bath} goes
through many distinct states, any small subsystem reaches equilibrium}. \end{quote} { Furthermore, we use our
proof of this (and the earlier indented statement) to show that equilibration occurs in initial product states
of the subsystem and bath, for almost all initial states of the bath. }

The notion of evolving through many different distinct states is mathematically encapsulated by the effective
dimension of the time-averaged state, $d^{\eff}(\omega)$ where $\omega = \av{\rho(t)}_t$. The connection between
this and the number of energy eigenstates is easily seen by expanding $\ket{\Psi(t)}$ (setting $\hbar=1$ for
convenience) as
\begin{equation}
  \ket{\Psi(t)} = \sum_k c_k e^{ - i E_k t} \ket{E_k},
\end{equation}
where $\sum_k |c_k|^2 =1$, and hence
\begin{equation}
  \rho(t) =\sum_{k,l} c_k c_l^{*} e^{ -i \left(E_k - E_l \right) t } \ketbra{E_k}{E_l}.
  \label{eqn:expansion1}
\end{equation}
From the non-degeneracy of the energy levels (implied by non-degenerate energy gaps),
\begin{equation}
  \omega = \av{\rho(t)}_t = \sum_k |c_k|^2 \proj{E_k},
  \label{eqn:expansion2}
\end{equation}
giving
\begin{equation}
  d^{\eff}(\omega) = \frac{1}{\tr(\omega^2)} = \frac{1}{\sum_k |c_k|^4}.\label{effective_dimension_omega}
\end{equation}

Similarly, the notion of evolving through many different distinct states of the bath is mathematically
encapsulated by the effective dimension of the time-averaged state of the bath, $d^{\eff}(\omega_B)$ where
$\omega_B = \av{\rho_B(t)}_t$. Note that in general, we expect the time-evolved state to explore a significant
portion of the bath state space, which is much larger than the subsystem's state space. Hence we anticipate that
$d^{\eff}(\omega_B)$ will be much larger than $d_S$.

{ We are now in the position to formulate our first theorem in rigorous mathematical terms. A quantity central
to this result is $D(\rho_S(t), \omega_S )$, the distance between $\rho_S(t)$, the state of the subsystem at
time $t$, and its time average, $\omega_S = \av{\rho_S(t)}_t $. In general $\rho_S(t)$ fluctuates around its
time average $\omega_S$ and the distance between them changes over time. To characterize these fluctuations we
look at the time average of the distance $\av{ D( \rho_S(t), \omega_S )}_t$. When this average is small, the
subsystem must spend almost all of its time very close to $\omega_S$. In other words, when $\av{ D( \rho_S(t),
\omega_S )}_t$ is small, the system equilibrates to $\omega_S$.


\bigskip
\noindent
{\bf Theorem 1~}
{\it Consider any state $\ket{\Psi(t)} \in \mathcal{H}$ evolving under a Hamiltonian with
non-degenerate energy gaps. Then
the average distance between} $\rho_S(t)$ {\it and its time average} $\omega_S$ {\it is bounded by}
\begin{equation}
    \av{ D( \rho_S(t), \omega_S )}_t
       \leq \frac{1}{2} \sqrt{\frac{d_S}{d^{\eff} (\omega_B)}}
       \leq \frac{1}{2} \sqrt{\frac{d_S^2}{d^{\eff} (\omega)}}.
\end{equation}
The proof of this result is given in appendix~\ref{app:2}.

By bounding $\av{ D( \rho_S(t), \omega_S )}_t$, our theorem tells us that the subsystem will equilibrate
whenever the effective dimension explored by the bath $d^{\eff} (\omega_B)$ is much larger than the subsystem
dimension $d_S$, or whenever the effective dimension explored by the total state $d^{\eff} (\omega)$ is much
larger than two copies of the subsystem (dimension $d_S^2$). In other words, if $\av{ D( \rho_S(t), \omega_S
)}_t$ is small, the system spends almost all its time close to the equilibrium state. Indeed, as distances are
always positive, it is easy to see that the proportion of the time for which $ D( \rho_S(t), \omega_S )$ is more
than $K$ times $\av{ D( \rho_S(t), \omega_S )}_t$ (with $K$ an arbitrary positive constant) must be less than
$1/K$. If we assume that the energy eigenvalues of $H$ have no rational dependencies (which is much stronger
than the non-degenerate energy gaps condition, but still one that  holds for generic perturbations of the
Hamiltonian) we can improve the bound on the proportion of time the subsystem spends away from equilibrium to
one exponential in $d^{\eff}(\omega)$ (see appendix~\ref{app:1}, Theorem~4).

We want to emphasize that this result about equilibration is completely general. That is, we did not assume
anything special about the interaction (apart from not having degenerate energy gaps - which rules out only a
set of Hamiltonians of measure zero), neither have we assumed any special properties of the bath. In particular,
we did not assume that the bath is characterized by some temperature $T$; in fact we did not assume that the
bath is in any kind of equilibrium at all! Furthermore, we did not make the rather standard assumption that the
system has a limited spread of energies $\Delta E$. Finally, we also did not make any assumptions about the form
of the equilibrium state $\omega_S$ of the subsystem; in particular there is no need for this state to have any
of the usual thermal properties, e.g.~to be of Boltzmannian form or similar. Indeed, since the bath is
completely arbitrary, so is the equilibrium state of the system. In other words, the equilibration phenomenon
that we describe is a general phenomenon, and needs not have any ``thermal'' aspects at all.

We also note that the bounds are completely independent from the energy eigenvalues. Indeed, the energy
eigenvalues did not play any role in establishing the bounds - they just gave rise to phases which averaged to
zero when taking the time average. This is a very important feature of our results. In fact, apart from the
discussion in the second half of section~\ref{sec:init-indep} energy does not appear anymore. Energy obviously plays a central
role in establishing the rate of time evolution and its exact details, but, as we have shown here, the long-time
behavior is characterized by bounds that are independent of energy.

Furthermore, it is not only the energy eigenvalues that play no role in our bounds - the form of the energy
eigenstates is also irrelevant. Indeed, the quantity $\frac{1}{2} \sqrt{\frac{d_S^2}{d^{\eff} (\omega)}} $ which
appears in our theorem is independent of the form of the energy eigenstates.

Finally, in the discussions above we referred to a specific partition of our total system into a "subsystem" and
its corresponding "bath". However, we note that the greater upper bound in Theorem 1 depends only on the
dimension of the subsystem, not on what that particular subsystem is, and on the effective dimension of $\omega$
which is independent of the way we partitioned our total system. Hence when one particular subsystem
equilibrates then {\it all} other subsystems of the same dimension equilibrate as well.

\subsection*{Equilibration of typical states}
Our theorem described above puts a bound on the fluctuation of the state of a subsystem around the time-average.
The next question is to find the cases in which the fluctuations are small so that the subsystem equilibrates.
As we will now show, almost all quantum states have this property. Intuitively the idea is the following.

Consider a Hilbert subspace $\mathcal{H}_R$ of the total Hilbert space of large dimension $d_R$. Consider now an
arbitrary orthogonal basis in this subspace. First, since the dimension of the subspace is large, it is
impossible to construct all basis vectors from only a few energy eigenstates. Thus, generically, each basis
vector is a superposition of many energy eigenstates. Second a typical state $|\Psi\rangle$ from this subspace
will have roughly equal overlap with each of the basis vectors. Hence a typical state $|\Psi\rangle$ will have a
significant overlap with many energy eigenstates, making $d^{\eff}(\omega)$ very large. Consequently, when we
chose a typical state from a large dimensional subspace $\mathcal{H}_R$ we find that the state is such that all
the small-dimension subsystems equilibrate.

More precisely, we can prove the following.

\bigskip
\noindent {\bf Theorem 2~~}
i) {\it The average effective dimension} $\left\langle d^{\eff} (\omega)  \right\rangle_{\Psi}$  {\it
where the average is computed over uniformly random pure states} $\ket{\Psi } \in {\cal H}_R \subset {\cal H}$
{\it is such that}

\begin{equation}
    \left\langle d^{\eff} (\omega)  \right\rangle_{\Psi} \geq \frac{d_R}{2}.\label{result2i}
  \end{equation}

\noindent ii) {\it For a random state} $\ket{\Psi} \in {\cal H}_R \subset {\cal H}$, {\it the probability}
${\Pr}_{\Psi} \left\{ d^{\rm eff}(\omega) < \frac{d_R}{4} \right\}$ that $d^{\rm eff}(\omega)$ is
smaller than $\frac{d_R}{4}$ {\it is exponentially small, namely}
\begin{equation}
    {\Pr}_{\Psi} \left\{ d^{\rm eff}(\omega) < \frac{d_R}{4} \right\}
                                               \leq 2 \exp\left( -c\sqrt{d_R} \right),\label{result2ii}
  \end{equation}
 {\it with a constant $c =  \frac{(\ln 2)^2}{72 \pi^3} \approx 10^{-4}$.}

\bigskip
\noindent The proof of (i) is given in appendix~\ref{app:2}, and the proof of (ii), which makes
use of Levy's lemma, is given in appendix~\ref{app:3}.

Point (i) essentially tells us that the average effective dimension is larger than half the dimension of the
Hilbert subspace, so when we draw states from a subspace of large dimension, the effective dimension $d^{\rm
eff}(\omega)$ of a typical state is large. Point (ii) makes the result even sharper, telling us that the
probability of having a small effective dimension is exponentially small.

We now apply these results to two specific cases.

\bigskip\noindent
{\bf Equilibration of generic states.}
We first address the question of what happens to a generic state of the whole system. In
other words, what happens to a state chosen at random from the total Hilbert space $\mathcal{H}$. The answer to
this question can be found immediately from the above results by taking the subspace
$\mathcal{H}_R$ to be the entire space $\mathcal{H}$, meaning  that in eqs. (\ref{result2i}) and
(\ref{result2ii}) we replace $d_R$ by $d$. Hence, for systems living in high dimensional Hilbert spaces, from
(\ref{result2i}) and (\ref{result2ii}) we expect that the effective dimension $d^{\rm eff}(\omega)$ is of the
order of $d$. Indeed, eq.  (\ref{result2ii})  tells us that the effective dimension will be smaller than
${d\over 4}$ only in an exponential small number of cases. Consequently, since $d=d_S d_B$, the factor
${{d_S^2}/{d^{\eff}(\omega)}}$ that governs the amount of fluctuations will be approximately equal to
${{d_S^2}/{d}}={d_S/{d_B}}$. In the case of systems composed of large numbers of particles, the dimension of the
Hilbert spaces of the system and bath grow exponentially with their number of particles. Hence this ratio will
drop off exponentially with the total number of particles whenever the number of particles in the subsystem is
no more than a constant fraction of the number in the bath, and thus each such subsystem equilibrates.

\bigskip\noindent
{\bf Equilibration of systems far from equilibrium.}
In the previous subsection we have shown that a typical state $|\Psi\rangle$ is such that any small enough
subsystem equilibrates. This however is not the end of the story. Indeed,  one can raise the following
legitimate question. Is it the case that states in which a subsystem is initially far from equilibrium will
equilibrate? The point is that states in which a subsystem is far from equilibrium are quite rare in the Hilbert
space so they are by no means "typical". Indeed, it can be shown that the overwhelming majority of the states in
the total Hilbert space are such that every small subsystem is already in a canonical
state~\cite{PopescuShortWinter}. In this section we focus specifically on what happens when the initial state of
the subsystem is far from equilibrium, and show that it too equilibrates.

We now consider the usual situation in which the question of equilibration is discussed. There is a bath,
consisting of a large number of particles, about which we know only some macroscopic parameters (such as
temperature), into which we place a much smaller subsystem. The initial state of the subsystem is arbitrary, but
uncorrelated with that of the bath. The question is, does the subsystem equilibrate?

In this section we prove that for any initial state of the subsystem, and almost all initial states of the bath,
the subsystem equilibrates. Note that this includes cases in which the subsystem is initially far from
equilibrium.

In our formalism, we describe the above situation as follows. We consider initial state of the full system (i.e.
of the subsystem and bath) to be a product state $\ket{\Psi}_{SB} = \ket{\psi}_S \ket{\phi}_B$. The initial
state of the system $\ket{\psi}_S$ is an arbitrary state in $\mathcal{H}_S$. We model the macroscopic parameters
of the bath by restricting its initial state $\ket{\phi}_B$ to lie within a particular subspace $\mathcal{H}^R_B
\subseteq \mathcal{H}_B$ (of dimension $d_B^R$). Note, however, that although the motivation for imposing a
restriction on the bath's Hilbert space was to model macroscopic parameters, our results apply to any restricted
Hilbert space $\mathcal{H}^R_B$. In particular it need not have any thermodynamic or macroscopic meaning.
Furthermore, $\mathcal{H}^R_B$ is only a restriction at the initial time, and the state of the bath may evolve
outside this space over time.

Given this setup, we can apply  our second theorem, eqs. (\ref{result2i}) and (\ref{result2ii})  with
$\mathcal{H}_R = \ket{\psi}_S \otimes {\cal H}_B^R$, and hence $d_R = d_B^R$. This gives $d^{\eff}(\omega) \geq
d_B^R/4$ for almost all initial states of the bath, and any initial state of the system.
 Hence given an initial product state, the subsystem will equilibrate for
any initial state of the system and almost any initial state of the bath (within $\mathcal{H}_B^R$), as long as
$d^R_B \gg d_S^2$.

The actual mechanism by which the subsystem equilibrates when it is put in contact with the bath is however
highly nontrivial. Indeed, as we show below, generically the state of the bath evolves through many distinct
states and does not reach equilibrium. Moreover it may move out of the initial restricted subspace
$\mathcal{H}^R_B$. At the same time, surprisingly, the subsystem that is in contact with it does equilibrate.
This is even more surprising, given the fact that we did not make any assumptions as to the form of the
interaction between the subsystem and the bath (apart from the Hamiltonian not having degenerate energy gaps).
Thus, in principle, it could be the case that the evolution of the subsystem is sensitive to the precise state
of the bath.

To show that the bath does not equilibrate generically, we will show that $d^{\eff}(\omega_B)$ the effective
dimension of the average state of the bath is much larger than $d^{\eff}(\rho_B(t))$, the effective dimension of
the bath state at any particular time, which essentially means that the bath state continues to evolve and does
not equilibrate to any particular state.  Indeed, note that because the two systems are in a pure entangled
state, $\textrm{rank}(\rho_B(t))=\textrm{rank}(\rho_S(t))\leq d_S$. Since the effective dimension of a state is
always less than its rank, we obtain that
\begin{equation}
d^{\eff}(\rho_B(t)) \leq d_S.
\end{equation}
In contrast, eq.~(\ref{eqn:deff_bounds}) implies that
\begin{equation}
  d^{\eff}(\omega_B) \geq d^{\eff}(\omega)/d_S.
\end{equation}
Hence for a generic state which, as discussed in Theorem 2 (ii) obeys $d^{\eff}(\omega)>{{d_R}\over4}$, we have
\begin{equation}
d^{\eff}(\omega_B) \geq \frac{d^R}{4 d_S}=\frac{d^R_B}{4 d_S} \gg d_S\geq d^{\eff}(\rho_B(t)).
\end{equation}

\section{Initial state independence}
\label{sec:init-indep} So far, we have addressed one issue associated with thermalisation, namely equilibration.
In particular we have shown that a small subsystem will equilibrate for almost all states drawn from a large
restricted subspace. We now address a second issue - how the equilibrium state reached by a subsystem depends on
the initial state. So far, each initial state obeying the restriction could cause the subsystem to equilibrate
to a different state. To represent this dependency we now explicitly denote the equilibrium state of the
subsystem by $\omega_S^{\Psi}$. However, when discussing thermalisation, we would usually  expect the
equilibrium state to depend only on macroscopic parameters (such as the temperature of the bath), and not on the
precise microscopic initial state. Our next result shows that this intuition caries over to our more general
setting, by proving that almost all states within a large restricted subspace lead to the same  equilibrium
state of a subsystem.

\bigskip
\noindent
{\bf Theorem 3~~}
i)  {\it Almost all initial states chosen from a large restricted subspace will yield
  the same equilibrium state of a small subsystem.  In particular,
  with $\langle \cdot \rangle_\Psi$  referring to the average over uniformly random pure states
  $\ket{\Psi(0)} \in {\cal H}_R \subset {\cal H_S \otimes H_B }$, and $\Omega_S = \av{\omega_S^{\Psi}}_{\Psi}$:
  \[
    \av{D(\omega_S^{\Psi}, \Omega_S )}_{\Psi} \leq \sqrt{\frac{d_S \, \delta}{4 d_R}} \leq \sqrt{\frac{d_S }{4 d_R}} .
  \]
The first inequality gives a tighter but more complicated bound, with
  \begin{equation}
  \delta = \sum_k \bra{E_k} \frac{\Pi_R}{d_R} \ket{E_k} \tr_S(\tr_B(\proj{E_k}))^2 \leq 1,
  \end{equation}
  where $\Pi_R$ is the projector onto $\mathcal{H}_R$.}

\medskip 
\noindent 
ii) {\it For a random state $\ket{\Psi} \in {\cal H}_R \subset {\cal H}$,  the probability
that $D(\omega_S^{\Psi}, \Omega_S)>\frac{1}{2} \sqrt{\frac{  d_S \delta}{d_R}} +\epsilon$  drops off
exponentially with $\epsilon^2 d_R$:}
\begin{equation}
  {\Pr}_{\Psi} \left\{ D(\omega_S^{\Psi}, \Omega_S)
                      > \frac{1}{2} \sqrt{\frac{d_S \delta}{d_R}} + \epsilon \right\}
        \leq 2 \exp\left( -c' \epsilon^2 d_R \right),
\end{equation}
 {\it with a constant  $c'= \frac{2}{9 \pi^3}$. Note that setting $\epsilon = d_R^{-1/3}$ yields a small average
  distance with very high probability when $d_R \gg d_S$.}

\bigskip
\noindent The proof is given in appendix~\ref{app:2} [part (i)] and~\ref{app:3} [part (ii)]. 
The quantity $\delta$ reflects the
average purity of the energy eigenstates on the subsystem  (with probability weights $\bra{E_k}
\frac{\Pi_R}{d_R} \ket{E_k}$ determined by their relevance to $\mathcal{H_R}$). The quantity $\delta$ lies in
the range $(\frac{1}{d_S}, 1)$, with highly entangled energy eigenstates yielding smaller
values.

\bigskip\noindent
{\bf Bath state independence.}
We now apply the result above to the situation described in the previous
section, in which the subsystem and bath are initially in the product state
$\ket{\Psi}_{SB} = \ket{\psi}_S \ket{\phi}_B$ in the restricted subspace $\mathcal{H}_R = \ket{\psi} \otimes
\mathcal{H}_R^B$. Previously, we have shown that generic states of the bath cause the subsystem to equilibrate.
However, the particular equilibrium state $\omega^{\Psi}_S$ of the subsystem could in principle depend on the
precise initial state of the bath $\ket{\phi}_B$. Here we show that this is not the case, and that almost all
states of the bath in $\mathcal{H}_R^B$ lead to the same equilibrium state of the subsystem.

The proof follows simply by applying Theorem 3 with $\mathcal{H}_R = \ket{\psi} \otimes \mathcal{H}_R^B$ and
hence $d_R = d_R^B$. From the weaker bound (not involving $\delta$) we see that almost all initial states of the
bath yield approximately the same equilibrium state $\omega_S$ of the subsystem, as long as $d_R^B \gg d_S$.
Note that since we do not use $\delta$, this result does not depend on any properties of the energy eigenstates.

\bigskip\noindent
{\bf Subsystem state independence.}
We now turn to the third aspect of thermalisation - subsystem state
independence. A cup of coffee left untouched in a closed room eventually reaches room temperature regardless of
its own initial temperature. In other words, a subsystem in contact with a bath reaches an equilibrium state
that only depends on the bath but not on its particular initial state. Could we derive this behavior in our
formalism? The issue of subsystem state independence turns out to be more complicated than the issues of
equilibration and bath independence and we have not been able to completely solve this problem, however we
present here a number of preliminary results and observations that we believe are crucial in understanding this
question.

There are many reasons why the issue of subsystem state independence is complicated. The first reason is that in
fact it is not always the case that the equilibrium state is independent of the initial state of the subsystem.
For a drastic example, consider the subsystem to be an atomic bomb. Then, depending on whether the bomb is
functional or not, the room temperature itself will be modified, and therefore also the final equilibrium state.
In other words, although the subsystem appears to be small, it may actually have a big impact on the bath, in
which case the equilibrium state depends on the initial state of the subsystem. Here it seems that not only the
dimension of the Hilbert spaces of the subsystem and bath are relevant (as in the previous sections), but the
value of energy itself. The energy however turns out not to be the end of the story, and perhaps not even the
key aspect, as we show below.

To start with, one can easily give an explicit example of a situation in which the equilibrium state of a
subsystem depends on its initial state regardless on how big the bath is, or on the energy scales involved. In
all our previous discussions the Hamiltonian was taken to be very general. The only condition we imposed on it
was that it has no degenerate energy gaps. This condition is sufficient to ensure that the subsystem interacts
with the bath ( i.e. $H\neq H_S+H_B$). However, this condition is not strong enough to ensure that there are no
conserved quantities of the subsystem. When there are such quantities, then clearly initial states of the
subsystem with different values of these quantities cannot equilibrate to the same state. An example would be a
Hamiltonian of the form
\begin{equation}
  \label{eq:diagonal-H}
  H = \sum_{nm} E_{nm} \proj{n}_S \otimes \proj{m}_B.
\end{equation}
As long as the energy eigenvalues $E_{nm}$ have no degenerate energy gaps (which could be achieved by choosing
them independently at random from some range), this is an interacting Hamiltonian. However, any operator $A =
\sum_n a_n \proj{n}_S $ on the subsystem will commute with $H$, and therefore is conserved. Evolution under this
Hamiltonian will dephase the state of the subsystem in the $\ket{n}$ basis, but cannot flip an $\ket{n}$ into an
$\ket{n'}$.

It is tempting to think that we could rule out such cases by only considering Hamiltonians that have no
conserved quantities on the subsystem. However, this is still not a sufficient condition, as illustrated by the
following counterexample (in which the system is a single spin, and the bath is composed of many spins).
\begin{equation}
  \label{eq:spin-bath-interaction}
  H = E \sigma_S^z + H_{\rm int} + H_B,
\end{equation}
where the eigenvalues of $H_{\rm int}$ and $H_B$ lie between $-1$ and $1$, and $E \gg 1$. It is easy to choose
an $H_{\rm int}$ such that there are no conserved quantities on the subsystem. In such cases, one might imagine
that in the long time limit it would be possible for all initial states of the subsystem to equilibrate to the
same state. However, it is easy to see that is not the case. Consider two initial states $\ket{\psi_+}_S
\ket{\phi}_B$ and $\ket{\psi_-}_S \ket{\phi}_B$ where $\sigma_z \ket{\psi_{\pm}} = \pm \ket{\psi_{\pm}} $. The
difference in expected total energy between these two states is close to $2E$ (the contribution of the terms
$H_{\rm int}$ and $H_B$ to the expected energy of any state lies between $+2$ and $-2$) and remains constant in
time. If the subsystem were to equilibrate to the same state in both cases, this difference in energy would have
to be accounted for by $H_{\rm int}$ and $H_B$. However these terms can generate an energy difference of at most
$4$. Hence such an equilibration process for the subsystem is impossible.

The above situation is indeed an abstract version of the ``atomic bomb'' example, in the sense that the energy
of the subsystem completely dominates that of the bath, despite the fact that the dimension of the Hilbert space
of the subsystem is much smaller than that of the bath.

It is important to note that a similar argument can be made for any globally conserved quantity in which the
total value is dominated by the contribution of the subsystem. Hence the issue is not one of energy. In fact we
believe that the only special role of the energy (more precisely, of the energy eigenvalues) is that of fixing
the time scale of the evolution. Indeed, in all our previous arguments the energy eigenvalues drop out of all
the calculations (as long as there are no degenerate energy gaps). Instead, we note that the ``atomic bomb''
example in its abstract form of the Hamiltonian in eq.~(\ref{eq:spin-bath-interaction}) is in fact an
approximation of the trivially non-thermalising Hamiltonian (\ref{eq:diagonal-H}), as it can be seen quite
easily that for large $E$ its eigenstates are arbitrarily close to product states.

The above discussion motivates us to consider the case in which the energy eigenstates are far from product. In
this situation we prove that almost all initial states of the subsystem lead to the same time-averaged state.
Hence when such states do equilibrate (i.e. spend almost all times close to their time average) they reach the
same equilibrium state.

We emphasize that as we mentioned in the introduction however, the issues of equilibration and subsystem
independence are in fact totally separated. Indeed, the question we address here is that of subsystem
independence of the {\it time-averaged} state of the subsystem, regardless on the fact whether or not the system
has only small fluctuations around this state, in which case the time-averaged state is an equilibrium state, or
not.

Consider the following situation: let the initial state of the system be a product state $\ket{\Psi} =
\ket{\psi}_S \ket{\phi}_B$, where now, in contrast to the previous section, $\ket{\phi}$ is fixed but
$\ket{\psi} \in {\cal H}_S$ is generic. I.e.~$\ket{\Psi}$ comes from a subspace ${\cal H}_R = {\cal H}_S \otimes
\ket{\phi}_B$. Again we apply Theorem 3. However, we now note that the weaker bound (not involving $\delta$) is
essentially useless, as $d_R=d_S$. Indeed, this is to be expected: the weaker bound does not take into account
any assumptions about the energy eigenstates  and we have argued above that there are cases in which the
time-averaged state of the subsystem depends strongly on its initial state.

This leaves us with the stronger bound involving $\delta$. Substituting $d_R=d_S$, we see that the time-averaged
state of the subsystem will be the same for almost all initial states as long as $\delta \ll 1$. However, this
is precisely the case we are interested in,in which the relevant energy eigenstates are far from product.

When the energy eigenstates are highly entangled, in the sense that $\delta \ll 1$, the equilibrium state of the
subsystem will therefore be approximately equal for almost all initial system states.

\section{Conclusions}

Understanding the basic laws of statistical mechanics from first principles is a holy grail of the subject. Here
we approach this question from a fundamentally different viewpoint than usual, in which the whole system is
described by a pure quantum state and the probabilistic behavior of a subsystem is an objective phenomenon, due
to quantum entanglement, rather than the result of subjective ignorance. We proved a general quantum result:
With almost full generality all interacting large quantum systems evolve in such a way that any small subsystem
equilibrates, that is, spends almost all time extremely close to a particular state. The only conditions we
require are that the Hamiltonian has no degenerate energy gaps (which rules out non-interacting Hamiltonians)
and that the state of the whole system contains sufficiently many energy eigenstates.

Virtually all physical situations satisfy these requirements. Firstly, all but a measure zero set of
Hamiltonians have non-degenerate energy gaps. Indeed, an infinitesimally small random perturbation will lift any
such degeneracy. Secondly, the vast majority of states in the Hilbert space are such that they contain (i.e.
have significant overlap with) very many energy eigenstates, as required in our proof. In particular, this
covers the physically interesting situation in which a subsystem, initially out of equilibrium, interacts with a
large bath. In this situation, we have proved that for every state of the subsystem and almost every state of
the bath, the subsystem equilibrates. Furthermore, we have also proved that the equilibrium state of the
subsystem is independent of the specific initial state of the bath and only depends on its macroscopic
parameters.

We would like to emphasize that the above two restrictions are the only conditions we require. We do not require
the assumptions usually made in statistical mechanics. For example the interactions could be strong and long
range and energy need not be an extensive quantity. Also the state of the bath could be a superposition of a
very large range of energy eigenstates and hence have no well defined temperature.

Although our original motivation was to study the phenomenon of thermalisation, we have found a much larger
range of phenomena: we discovered that reaching equilibrium is an almost universal behavior of large quantum
systems. Ordinary thermalization is just a particular example of this behavior; it occurs in specific situations
and it has a number of additional characteristics that we did not address in the present paper. For example, the
equilibrium state of a subsystem that is put into contact with a thermal bath, is largely independent of the
initial state of the subsystem. Also the equilibrium state has a particular form (generally Boltzmannian). These
issues are open questions for future work.

\bigskip
\noindent 
{\bf Note added.} After the completion of this work we became aware of an independent work by P.
Reimann published very recently, Phys. Rev. Lett. 101, 190403 (2008), which takes first steps towards solving
the thermalisation problem along similar lines as in the present paper.

\acknowledgments

We acknowledge support from the UK EPSRC ``QIP IRC'' grant and the EC project QAP. AJS also acknowledges support
from a Royal Society University Research Fellowship, and AW is supported by a U.K.~EPSRC Advanced Research
Fellowship and a Royal Society Wolfson Merit Award.

\appendix


\section{Proofs of expectation values}
\label{app:2}
In this appendix, we derive several expectation values that are used in the main paper.

\medskip
\begin{proof}[of Theorem 1]
We first relate the trace distance to a less natural, but more mathematically tractable distance measure (the
square of the Hilbert-Schmidt distance) using a standard bound~\cite{Fuchs-vandeGraaf}
\begin{equation}
  D(\rho_1, \rho_2) = \frac{1}{2}\tr_S \sqrt{ ( \rho_1 - \rho_2)^2}
                    \leq \frac{1}{2} \sqrt{d_S \tr_S ( \rho_1 - \rho_2)^2}.
\end{equation}
Using the concavity of the square-root function, we therefore have
\begin{equation}
\av{D(\rho_S(t), \omega_S )}_t \leq \sqrt{d_S \av{ \tr_S \left( \rho_S(t) - \omega_S \right)^2}_t}.
\end{equation}
Using the expansions in terms of the energy eigenstates given in eqs.~(\ref{eqn:expansion1}) and
(\ref{eqn:expansion2}),
\begin{equation}
  \rho_S(t) - \omega_S = \sum_{k \neq l } c_k c_l^{*} e^{-i(E_k - E_l)t} \tr_B \ketbra{E_k}{E_l},
\end{equation}
where $\sum_{k \neq l }$ is shorthand for a sum over $k$ and $l$ omitting all terms in which $k=l$. Hence,
\begin{align*}
  &\!\!\!\!\!\!\!\!\!
     \av{ \tr_S (\rho_S(t) - \omega_S)^2}_t \\
  &= \sum_{k \neq l } \sum_{m \neq n} \mathcal{T}_{klmn} \tr_S \bigl( \tr_B \ketbra{E_k}{E_l} \tr_B \ketbra{E_m}{E_n} \bigr),
\end{align*}
where
\begin{equation} \label{eqn:Eklmn}
   \mathcal{T}_{klmn} = c_k c_l^* c_m c_n^* \av{e^{- i (E_k -E_l + E_m - E_n) t}}_t.
\end{equation}
Evaluating this time average, using the fact that the Hamiltonian has non-degenerate energy gaps, and that the
sums only include those terms where $k \neq l$ and $m \neq n$, we find that the only non-zero terms are those
where $k=n$ and $l=m$, giving
\begin{equation}\begin{split}
  &\!\!\!\!\!\!\!\!
   \av{ \tr_S (\rho_S(t) - \omega_S)^2}_t \\
  &=\sum_{k \neq l } |c_k|^2 |c_l|^2 \tr_S \bigl( \tr_B \ketbra{E_k}{E_l} \tr_B \ketbra{E_l}{E_k} \bigr) \\
  &=\sum_{k \neq l } |c_k|^2 |c_l|^2 \sum_{ss'bb'} \braket{sb}{E_k}\braket{E_l}{s'b}\braket{s'b'}{E_l}\braket{E_k}{sb'} \\
  &=\sum_{k \neq l } |c_k|^2 |c_l|^2 \sum_{ss'bb'} \braket{sb}{E_k}\braket{E_k}{sb'}\braket{s'b'}{E_l}\braket{E_l}{s'b} \\
  &=\sum_{k \neq l } |c_k|^2 |c_l|^2 \tr_B \bigl( \tr_S \ketbra{E_k}{E_k} \tr_S \ketbra{E_l}{E_l} \bigr)\\
  &= \sum_{k \neq l }\tr_B \bigl( \tr_S ( |c_k|^2 \proj{E_k} )
                                   \tr_S ( |c_l|^2 \proj{E_l} ) \bigr) \\
  &= \tr_B \omega_B^2
      - \sum_k |c_k|^4 \tr_S \left(\left(\tr_B \proj{E_k} \right)^2 \right) 
   \leq \tr_B \omega_B^2,
\end{split}\end{equation}
where $\omega_B = \tr_S \omega$. To obtain a further bound, we invoke weak subadditivity of the R\'{e}nyi
entropy (see e.g.~\cite{vandam}),
\begin{equation}
  \label{eqn:deff_bounds}
  \tr (\omega^2) \geq \frac{\tr_B(\omega_B^2)}{\textrm{Rank}(\rho_S)}
                 \geq \frac{\tr_B (\omega_B^2)}{d_S},
\end{equation}
and therefore
\begin{equation}\begin{split}
  \av{D( \rho_S(t) , \omega_S)}_t &\leq \frac{1}{2} \sqrt{d_S \tr_B \left( \omega_B^2 \right) } \\
                                  &\leq \frac{1}{2} \sqrt{d_S^2 \tr \left( \omega^2 \right) }.
\end{split}\end{equation}
Using the definition of the effective dimension (\ref{eqn:notation2}), these are the
bounds as stated in the theorem.
\end{proof}

In order to prove Theorems 2(i) and 3(i), we require a couple of additional mathematical tools, that are also
used in~\cite{PopescuShortWinter}. The first is the simple and very helpful identity
\begin{equation}
\tr (AB) = \tr ((A \otimes B ) \swap ),
\end{equation}
where $\swap$ is the SWAP operator of the two systems.
This equation is easily proved by expanding it in a basis. The second result we employ is that $\av{ \proj{\Psi} \otimes \proj{\Psi} }_{\Psi}$ is proportional to the projector onto the symmetric subspace. This is actually the simplest instance of the representation theory of the unitary group. It  implies the following lemma:

\bigskip\noindent
{\bf Lemma~~} 
\emph{For states $\ket{\Psi}\in \mathcal{H}_R \subseteq \mathcal{H}$
\begin{equation}
  \av{ \proj{\Psi} \otimes \proj{\Psi} }_{\Psi} = \frac{ \Pi_{ RR} (\1 + \swap)}{d_R (d_R+1)},
\end{equation}
where $\1$ is the identity operator on $\mathcal{H} \otimes \mathcal{H}$, and 
$\Pi_{ RR} = \Pi_R \otimes \Pi_R$ is the projector onto $\mathcal{H}_R \otimes \mathcal{H}_R$.}
\qed

\bigskip\noindent
For conciseness, in the proofs of Theorems 2(i) and 3(i) we abbreviate energy eigenstates by their indices, such
that $\ket{E_k} \equiv \ket{k}$ etc. Similarly, we write $\ket{E_k} \otimes \ket{E_l}$ as $\ket{kl}$ etc. It is
also helpful to note that time-averaging a state corresponds to de-phasing it in the energy eigenbasis (due to
the non-degeneracy of energy levels). We denote this dephasing map by $ \$[\rho] := \sum_k \proj{k} \rho
\proj{k}$, such that $\omega = \av{\proj{\psi}}_t = \$[\proj{\psi}]$.

With these tools and notation in hand, we now prove the results needed for the main paper

\begin{proof}[of Theorem~2(i)]
We first prove a bound on the expected purity of $\omega$,
\begin{equation}\begin{split}
\av{\tr(\omega)^2}_{\Psi}
&= \av{ \tr((\omega \otimes \omega)\swap) }_{\Psi}  \\
&=\tr\left(\$ \otimes \$ \left[\av{ \proj{\Psi} \otimes \proj{\Psi}}_{\Psi}\right]\swap \right) \\
&=\tr\left(\$ \otimes \$ \left[\frac{ \Pi_{ RR} (\1 + \swap)}{d_R (d_R+1)} \right]\swap\right) \\
&= \sum_{kl} \tr\left(\proj{k l}  \left(\frac{ \Pi_{ RR}  (\1 + \swap)}{d_R (d_R+1)} \right) \ketbra{k l}{k l} \swap \right)  \\
&= \sum_{kl} \tr\left(\ketbra{k l}{l k} \right)  \left(\frac{ \bra{kl} \Pi_{ RR}  (\ket{kl}+ \ket{lk})}{d_R (d_R+1)} \right)  \\
&= \sum_{k}  \frac{ 2 \bra{kk} \Pi_{ RR} \ket{kk}}{d_R (d_R+1)} \\
&\leq \sum_{k}  \frac{ 2 \bra{k} \Pi_{ R} \ket{k}}{d_R (d_R+1)} 
 < \frac{2}{d_R}.
\end{split}\end{equation}
It then follows straightforwardly that
\begin{equation}
\av{d^{\eff}(\omega)}_{\Psi} = \av{\frac{1}{\tr(\omega)^2}}_{\Psi}  \geq \frac{1}{\av{\tr(\omega)^2}_{\Psi} }
> \frac{d_R}{2},
\end{equation}
concluding the proof.
\end{proof}

\begin{proof}[of Theorem~3(i)]
As in the proof of Theorem 1, we first relate the trace distance to
a less natural, but more mathematically tractable distance
measure~\cite{Fuchs-vandeGraaf},
\begin{eqnarray}
\av{D(\omega_S, \av{\omega_S}_{\Psi} ) }_{\Psi}  &\leq& \av{\frac{1}{2} \sqrt{d_S \tr \left(\omega_S - \av{\omega_S}_{\Psi} \right)^2} }_{\Psi} \nonumber \\
&\leq& \frac{1}{2} \sqrt{d_S\av{ \tr \left(\omega_S -
\av{\omega_S}_{\Psi} \right)^2} }_{\Psi}.
\label{eqn:distance-relation}
\end{eqnarray}
We then prove bounds on the averaged term
\begin{eqnarray}
&&\!\!\!\!\!\!\!
  \av{\tr \left(\omega_S - \av{\omega_S}_{\Psi} \right)^2 }_{\Psi} \nonumber \\
&&=\av{\tr \left(\omega_S^2\right)}_{\Psi}  - \tr_S\left(\av{\omega_S}_{\Psi}^2\right)  \nonumber \\
&&=\tr \left( \left(\av{ \omega_S \otimes \omega_S }_{\Psi} -  \av{ \omega_S }_{\Psi} \otimes \av{\omega_S }_{\Psi}   \right) \swap \right) \nonumber \\
&&=\tr_{SS}\!\left(\! \tr_{BB}\!\! \left(\$ \otimes \$\! \left[ \! \av{ \proj{\Psi}\! \otimes\! \proj{\Psi} }_{\Psi}\!  - \!\frac{\Pi_R}{d_R}\! \otimes \!\frac{\Pi_R}{d_R} \right] \right) \swap \right) \nonumber \\
&&=\tr_{SS} \left(\! \tr_{BB} \!\!\left( \$ \otimes \$ \left[ \frac{ \Pi_{ RR} (\1 + \swap)}{d_R (d_R+1)} - \frac{\Pi_{ RR}}{d_R^2} \right]  \right) \swap \right)\nonumber \\
&& \leq \tr_{SS}\! \left(\! \tr_{BB} \!\!\left( \$ \otimes \$ \left[ \frac{ \Pi_{ RR} \swap}{d_R^2} \right]  \right) \swap \right)\nonumber \\
&&=\sum_{kl} \tr_{SS}\! \left(\! \tr_{BB} \!\!\left( \proj{kl}  \frac{ \Pi_{ RR} }{d_R^2} \ketbra{lk}{kl} \right) \swap \right)\nonumber \\
&&=\sum_{kl} \frac{\bra{kl} \Pi_{ RR} \ket{lk}}{d_R^2} \tr_S\! \big( \tr_B \left(\proj{k} \right) \tr_B \left(\proj{l} \right)   \big)\nonumber \\
&&\leq \sum_{kl} \frac{\bra{k}\! \Pi_R \proj{l} \!\Pi_R \ket{k}}{ d_R^2} \tr_S\! \left( \frac{\left( \tr_B\proj{k} \right)^2+\left( \tr_B \proj{l} \right)^2 }{2}  \right)\nonumber \\
&&=\frac{1}{d_R} \sum_k  \bra{k}\frac{\Pi_R}{d_R} \ket{k}  \tr_S\! \left( \left( \tr_B\proj{k} \right) ^2  \right) \nonumber \\
&&\leq \frac{1}{d_R} \sum_k  \bra{k}\frac{\Pi_R}{d_R} \ket{k}  
  = \frac{1}{d_R}. 
\end{eqnarray}
Note that in the second inequality (9th line) we have used the fact that $\tr_S ( \tr_B \proj{k} - \tr_B
\proj{l})^2 \geq 0$ (it is the trace of a positive operator).

Inserting these results into eq.~(\ref{eqn:distance-relation}), specifically, the bounds given by the
third line from bottom and the last line, we obtain the inequalities stated in the theorem.
\end{proof}

\section{Uses of Levy's Lemma}
\label{app:3}
In this appendix we prove Theorems 2(ii) and 3(ii), in which
exponential bounds are placed on the proportion of states which have
values of $d^{\eff}(\omega)$ and $D(\omega_S^{\Psi}, \Omega_S)$ far
from the average value, using Levy's Lemma for measure concentration
on a hypersphere.

\bigskip\noindent
{\bf Levy's Lemma~\cite{MilmanSchechtman}~~}
\emph{Let $f:S^{D-1} \rightarrow \RR$ be a real-valued function
  on the $(D-1)$-dimensional Euclidean sphere (which we think of as embedded into $D$-dimensional
  Euclidean space), with Lipschitz constant
  $\lambda = \sup_{x_1, x_2} |f(x_1) - f(x_2)| / |x_1 - x_2|_2$. Then,
  for a uniformly random point $X \in S^{D-1}$,
  \begin{equation}
    {\Pr}_X \bigl\{ f(X) > \langle f \rangle + \epsilon \bigr\}
        \leq 2 \exp\left( - \frac{ D \epsilon^2}{9 \pi^3 \lambda^2} \right).
        \qquad \squareforqed
  \end{equation}}

\bigskip\noindent
Note that the pure quantum states $\ket{\Psi} \in \mathcal{H}_R$ can
be thought of as lying on a $(2 d_R-1)$-dimensional hypersphere, with
coordinates given by the real and imaginary components of the
state's overlap with an orthonormal basis $\ket{\hat{n}}$ of
$\mathcal{H}_R$:
\begin{equation}
x_{2n-1}(\Psi)= {\rm{Re}}[\braket{\hat{n}}{\Psi}], \quad x_{2n}(\Psi)= {\rm{ Im}}[\braket{\hat{n}}{\Psi}].
\end{equation}
In this coordinate system Euclidean and Hilbert space norm coincide:
$| \vec{x}(\Psi_1) - \vec{x}(\Psi_2)|_2 = \bigl|\ket{\Psi_1} - \ket{\Psi_2}\bigr|_2$.

\medskip
Because it involves the simplest application of Levy's Lemma, we
begin with the proof relating to initial state independence.

\bigskip
\begin{proof}[of Theorem~3(ii)] To prove that almost all initial states $\ket{\Psi}$
yield the same equilibrium state, we apply Levy's Lemma directly to
the function $f(\Psi) \equiv f(\vec{x}(\Psi)) = D(\omega_S^{\Psi},
\Omega_S)$ on the $(2 d_R-1)$-dimensional hypersphere of quantum
states. Using an analogue of eq.~(\ref{eqn:lipschitz}) to give
\begin{equation}
| D(\omega_S^{\Psi_1}, \Omega_S) - D(\omega_S^{\Psi_2}, \Omega_S)| 
   \leq \bigl|\ket{\Psi_1} - \ket{\Psi_2}\bigr|_2,
\end{equation}
we find that the Lipschitz constant of the function satisfies
$\lambda \leq 1$. Substituting this into Levy's Lemma, as well as
the average value obtained in Theorem 3(i), we obtain the desired
result.
\end{proof}
\bigskip
\begin{proof}[of Theorem 2(ii)] Unfortunately we cannot prove the
result in Theorem 2(ii) directly by applying Levy's Lemma to
$d^{\eff}(\omega)$. Instead, we apply it to the function
\begin{equation}
f(\Psi) \equiv f(\vec{x}(\Psi)) = \ln \left( \tr \left(\tilde{\$}[\proj{\Psi}]^2 \right)\right),
\end{equation}
where the super-operator $\tilde{\$}$ acts on the subspace
$\mathcal{H}_T \subseteq \mathcal{H}$ spanned by energy eigenstates
with nonzero projection onto $\mathcal{H}_R$ (i.e. states $\ket{k}$
satisfying $\bra{k} \Pi_R \ket{k} \neq 0$). The subspace
$\mathcal{H}_T$ contains all states which could arise during the
evolution of an initial state in $\mathcal{H}_R$, and $\tilde{\$}$
maps these states back into $\mathcal{H}_R$ according to
\begin{equation}
\tilde{\$}[\rho ] = \sum_k \ketbra{\tilde{k}}{k} \rho \ketbra{k}{\tilde{k}} \quad {\rm and} \quad
\ket{\tilde{k}} = \frac{1}{\sqrt{\bra{k} \Pi_R \ket{k}}}\Pi_R \ket{k}.
\end{equation}
Note that  when the Hamiltonian commutes with $\Pi_R$, $\tilde{\$}$
is identical to the normal $\$$ on $\mathcal{H}_T$. Computing the
average value of our function we find
\begin{equation}\begin{split}
&\!\!\!\!\!
 \av{\ln  \left( \tr \left( \tilde{\$}[\proj{\Psi}]^2
\right)\right)}_{\Psi}  \\
&\leq \ln \av{ \tr \left(
\tilde{\$}[\proj{\Psi}]^2 \right)}_{\Psi} \\
& =\ln  \tr \left( \tilde{\$} \otimes \tilde{\$}\left[\av{\proj{\Psi}
\otimes \proj{\Psi}}_{\Psi}\right] \swap \right)  \\
&=\ln \left( \tr \left( \tilde{\$} \otimes
\tilde{\$}\left[\frac{\Pi_{RR}(\1 + \swap)}{d_R(d_R+1)}\right] \swap
\right)
\right) \\
&=\ln \left(\sum_{kl} \tr\left(\ketbra{\tilde{k}\tilde{l}}{k l}
\left(\frac{ \Pi_{ RR}
(\1 + \swap)}{d_R (d_R+1)} \right) \ketbra{k l}{\tilde{l}\tilde{k}}  \right) \right)  \\
&=\ln \left( \sum_{kl}
\braket{\tilde{l}\tilde{k}}{\tilde{k}\tilde{l}} \left(\frac{
\bra{kl} \Pi_{ RR} (\ket{kl}+ \ket{lk})}{d_R (d_R+1)} \right)\right)
\\
&\leq \ln \left( \sum_{kl}
\braket{\tilde{l}\tilde{k}}{\tilde{k}\tilde{l}} \left(\frac{
\bra{kl} \Pi_{ RR} (\ket{kl}+ \ket{lk})}{d_R (d_R+1)}
\right)\right)\\
&\leq \ln \left( \frac{2}{d_R(d_R+1)} \sum_{kl} \bra{l k }\Pi_{RR} \ket{k l} \right) \\
&=\ln \left(  \frac{2}{d_R(d_R+1)} \sum_{k} \bra{k }\Pi_{R} \ket{k} \right)
 < \ln \left( \frac{2}{d_R} \right).
\end{split}
\end{equation}
To bound the Lipschitz constant of the function $f(\Psi)$, we actually employ a further function
\begin{equation}
g(\Psi)  = \ln  \tr \left[ \left( \sum_{n} \proj{\hat{n}}
\tilde{\$}[\proj{\Psi}] \proj{\hat{n}} \right)^2 \right],
\end{equation}
where $\ket{\hat{n}}$ is an orthonormal basis of $\mathcal{H}_R$.
Writing \begin{equation} t_{n k 0} = {\rm
Re}[\braket{\hat{n}}{\tilde{k}} \braket{k}{\Psi}] \quad {\rm and}
\quad t_{n k 1} = {\rm Im}[\braket{\hat{n}}{\tilde{k}}
\braket{k}{\Psi}]
\end{equation}
it follows that
\begin{equation}
g(\Psi) = \ln \tr \left[ \left( \sum_{n k z} t_{n kz}^2 \proj{\hat{n}} \right)^2 \right] =\ln \sum_{n} \left(
\sum_{kz} t_{n kz}^2 \right)^2. \end{equation} To bound the Lipschitz constant of $g$, it is sufficient to find
an upper bound on its gradient.
  \begin{equation}
    \frac{\partial g}{\partial t_{n kz}}
            = \frac{1}{\sum_{n'} \left( \sum_{k'z'} t_{n 'k'z'}^2 \right)^2}
                                           .2.2.t_{n kz} \sum_{k'z'} t_{n
                                           k'z'}^2.
  \end{equation}
  Introducing the notation $p_{n} = \sum_{kz} t_{n kz}^2$, and noting that $\sum_{n} p_{n}
  =1$, we find
  \begin{equation} \begin{split}
    | \nabla g |^2 = \sum_{n k z} \left( \frac{\partial g}{\partial t_{n kz}} \right)^2&= \frac{16 \sum_{n} p_{n}^3}{\left( \sum_{n} p_{n}^2
    \right)^2} \\
                                 &\leq \frac{16 \left( \sum_{n} p_{n}^2 \right)^{3/2}}{\left( \sum_{n} p_{n}^2 \right)^2} \\
                                 &=\frac{16}{\left(\sum_{n} p_{n}^2\right)^{1/2}} \\
                                 &\leq 16 \sqrt{d_R},
  \end{split} \end{equation}
  and hence that the Lipschitz constant of $g$ is at most
  $4\sqrt[4]{d_A}$.

To obtain the Lipschitz constant of $f$, we note that $g(\Psi) \geq
f(\Psi)$, with equality if $\{ \ket{\hat{n}} \}$ is an eigenbasis of
$\tr_B \proj{\Psi}$. Now, for any two vectors, we may without loss
of generality assume that $f(\Psi_1) \geq f(\Psi_2)$, and take $\{
\ket{\hat{n}} \}$ to be the eigenbasis of $\tr_B \proj{\Psi_2}$.
Thus,
  \begin{equation}
    f(\Psi_1)-f(\Psi_2) \leq g(\Psi_1)-g(\Psi_2) \leq 4\sqrt[4]{d_A} \bigl| \ket{\Psi_1}-\ket{\Psi_2} \bigr|_2,
  \end{equation}
and hence the  Lipschitz constant of $f$ is also upper-bounded by
$4\sqrt[4]{d_A}$.

Applying Levy's Lemma to $f(\Psi)$, and using the observation that
${\Pr}\{x>a\}\leq b$ and $x \geq y$ implies ${\Pr}\{y>a\}\leq b$ to
substitute the bound on $\av{f(\Psi)}_{\Psi}$ obtained above, and
the further bound
\begin{equation}
\ln \left( \tr \left( \tilde{\$}[\proj{\Psi}]^2 \right)\right) \geq
\ln \left( \tr \left( \$[\proj{\Psi}]^2 \right)\right)
\end{equation}
gives
\begin{equation}
{\Pr}_{\Psi} \left\{\ln \left( \tr \left( \$[\proj{\Psi}]^2
\right)\right) > \ln \frac{2 e^{\epsilon}}{d_R}  \right\}
      \leq 2 \exp\left( - \frac{\epsilon^2\sqrt{d_R}}{72 \pi^3}
      \right).
\end{equation}
Negating both sides of the expression inside the braces, and then
taking their exponent yields
\begin{equation}
{\Pr}_{\Psi} \left\{d^{\eff}(\omega) < \frac{d_R}{2 e^{\epsilon}}
\right\}
      \leq 2 \exp\left( - \frac{\epsilon^2\sqrt{d_R}}{72 \pi^3}
      \right).
\end{equation}
Finally setting $\epsilon=\ln 2$ yields the desired result.
\end{proof}

\section{Fluctuations from the time-average}
\label{app:1}
Here we prove an exponential bound on the proportion of time the state spends a long way from its
equilibrium state, using measure concentration results~\cite{ledoux:ams}, and a stronger assumption on $H$.

Making the assumption that the eigenenergies $E$ of $H$ have no rational dependencies (which is much stronger
than our non-degenerate energy gaps condition), the trajectory
\begin{equation}
  \ket{\Psi(t)} = \sum_k e^{-itE_k}c_k\ket{E_k}
\end{equation}
over time fills the torus
\begin{equation}
  {\cal T} = \left\{ \Psi(\underline{\a}) := \sum_k e^{i\a_k} c_k \ket{E_k} : 0\leq \a_k \leq 2\pi \right\}
\end{equation}
\emph{uniformly}. What this means in fact is that an ergodic theorem holds, equating the time averages of all
continuous functions of the state with their respective ``ensemble'' averages, i.e.~with integration over
independent phase angles $\a_k$. The same is true for the indicator functions of open sets, as the one in the
following theorem.

\bigskip\noindent
{\bf Theorem 4~~} {\it
  \label{thm:levy-torus}
  Under the above assumption of ergodicity,
  \begin{equation}\begin{split}
    {{\Pr}}_t&\left\{ D(\rho_S(t),\omega_S) > \sqrt{\frac{d_S}{d^{\rm eff}(\omega_B)}} + \epsilon \right\} \\
           &\phantom{===}
            = {{\Pr}}_{\underline{\a}}\left\{ D(\rho_S(\underline{\a}),\omega_S)
                                             > \sqrt{\frac{d_S}{d^{\rm eff}(\omega_B)}} + \epsilon \right\} \\
           &\phantom{===}
            \leq \exp\left( -c'' \epsilon^4 d^{\rm eff}(\omega)\right),
  \end{split}\end{equation}
  with a  constant $c'' = \frac{1}{128\pi^2}$ }
  \bigskip

\begin{proof}
  We already argued the first equality, so only the second
  inequality is to be proved.

  From Theorem 1 in the main paper, the average of $D(\rho_S(\underline{\a}),\omega_S)$ is
  upper bounded by $\frac{1}{2}\sqrt{\frac{d_S}{d^{\eff} (\omega_B)}}$, hence
  by Markov's inequality its median is upper bounded by
  $\sqrt{\frac{d_S}{d^{\rm eff} (\omega_B)}}$.

  The space on which we want to study the measure concentration is the
  direct product of unit circles. As the metric on that space we choose
  the following weighted $\ell^1$ metric:
  \begin{equation}
    d(\underline{\a},\underline{\b}) = \frac{1}{2\pi} \sum_k |c_k|^2 |\a_k-\b_k|.
  \end{equation}
  The $k$-th factor has diameter $|c_k|^2$, so using~\cite[Thm.~4.4]{ledoux:ams},
  we conclude that the concentration function $\vartheta$ of this space obeys
  \begin{equation}
    \vartheta(r) \leq e^{-r^2/8L^2}, \text{ where } L^2 = \sum_k |c_k|^4 = 1/d^{\rm eff}(\omega).
  \end{equation}
  ``Concentration function'' means that for any set $A \subset [0;2\pi)^d$ of
  measure $\geq 1/2$, the $r$-neighbourhood
  $A_r = \{ \underline{\a} \text{ s.t. }
             \exists \underline{\b}\in A\ d(\underline{\a},\underline{\b}) \leq r \}$
  has measure at least $1-\vartheta(r)$.

  Hence, all that remains is to relate the distance $r$ to the variation
  of the function $D(\rho_S(\underline{\a}),\omega_S)$. We first obtain the general result
  \begin{equation}\begin{split} \label{eqn:lipschitz}
    \bigl| D(\rho_S(\underline{\a}),\omega_S) - D(\rho_S(\underline{\b}),\omega_S) \bigr|
                  &
                   \leq D(\rho_S(\underline{\a}), \rho_S(\underline{\b}) ) \\
                  &\!\!\!\!\!\!\!\!\!\!
                   \leq D(\rho(\underline{\a}), \rho(\underline{\b}))\\
                  &\!\!\!\!\!\!\!\!\!\!
                   = \sqrt{1 - |\bra{\psi(\underline{\a})}\psi(\underline{\b})\rangle|^2} \\
                  &\!\!\!\!\!\!\!\!\!\!
                   \leq \left| \ket{\psi(\underline{\a})} - \ket{\psi(\underline{\b})} \right|_2.
  \end{split}\end{equation}
  Together with
  \begin{equation}\begin{split}
  \left| \ket{\psi(\underline{\a})} - \ket{\psi(\underline{\b})} \right|_2^2
                  & = \sum_k |c_k|^2 \left| e^{i\a_k} - e^{i\b_k} \right|^2 \\
                  &\leq  2 \sum_k |c_k|^2 \left| e^{i\a_k} - e^{i\b_k} \right| \\
                  &\leq 2\sum_k |c_k|^2 |\a_k-\b_k|
                      = 4\pi d(\underline{\a},\underline{\b}),
  \end{split}\end{equation}
  this results in
  \begin{equation}
    \bigl| D(\rho_S(\underline{\a}),\omega_S) - D(\rho_S(\underline{\b}),\omega_S) \bigr|
                                         \leq \sqrt{4\pi}\sqrt{d(\underline{\a},\underline{\b})}.
  \end{equation}
  So, to change the value of $D(\rho_S,\omega_S)$ by more than $\epsilon$
  above the median, one has to go from an element in the submedian set
  a distance of at least $r=\frac{\epsilon^2}{4\pi}$ in our chosen metric $d$.
  Inserting this into the above concentration function
  yields the result.
\end{proof}

\end{document}